\documentclass[twocolumn,nonbibnotes,nofootinbib,notitlepage,longbibliography]{revtex4-2}
\usepackage{amsmath}
\usepackage{amssymb}
\usepackage{bm}
\usepackage{epsfig}
\usepackage{graphicx}
\usepackage{color}
\usepackage[colorlinks=true,citecolor=blue,urlcolor=black]{hyperref}
\usepackage[english]{babel}

\usepackage[bbgreekl]{mathbbol}

\newcommand{\add}[1]{\textcolor{black}{#1}}

\begin{document}

\title{Valley polarization fluctuations, bistability, and switching in two-dimensional semiconductors}

\author{M.A. Semina, M.M. Glazov}
\affiliation{Ioffe Institute, 26 Polytechnicheskaya, 194021 St. Petersburg, Russia}

\author{C. Robert, L. Lombez, T. Amand, X. Marie}
\affiliation{Universit\'e de Toulouse, INSA-CNRS-UPS, LPCNO, 135 Avenue Rangueil, 31077 Toulouse, France}

\date{\today}

\begin{abstract}
We study theoretically nonlinear valley polarization dynamics of excitons in atom-thin semiconductors. The presence of significant polarization slows down valley relaxation due to an effective magnetic field resulting from exciton-exciton interactions. We address temporal dynamics of valley polarized excitons and study the steady states of the polarized exciton gas. We demonstrate bistability of the valley polarization where two steady states with low and high valley polarization are formed. We study the effects of fluctuations and noise in such system. We evaluate valley polarization autocorrelation functions and demonstrate that for a high-polarization regime the fluctuations are characterized by high amplitude and long relaxation time. We study the switching between the low- and high-valley polarized states caused by the noise in the system and demonstrate that the state with high valley polarization is preferential in a wide range of pumping rates.
\end{abstract}

\maketitle

\section{Introduction}

Spin and valley physics of electrons and excitons in atomically-thin crystals is a vibrant field of experimental and theoretical research~\cite{Yang:2015aa,Dufferwiel:2017aa,PhysRevLett.119.137401,Molina-Sanchez:2017aa,app8071157,PhysRevResearch.1.022007,Shinokita:2019aa,Tang:2019ab,PhysRevB.101.115307,PhysRevResearch.2.023322,Lloyd:2021wk,Jiang:spin,PhysRevMaterials.5.044001,Kravtsov_2021,Glazov_2021} owing to the prospects of applications in quantum technologies and intriguing physics of spin-orbit and exchange interactions~\cite{ivchenko05a,dyakonov_book,glazov2018electron}. 

Fast spin and valley depolarization of bright excitons in monolayer transition metal dichalcogenides is mainly controlled by the long-range exchange interaction between an electron and a hole forming an exciton~\cite{glazov2014exciton,Yu:2014fk-1,PhysRevB.89.205303,PhysRevB.90.161302,PSSB:PSSB201552211,prazdnichnykh2020control} similarly to the case of quasi-two-dimensional excitons in  semiconductor quantum well structures~\cite{maialle93,goupalov98,ivchenko05a}. The long-range exchange interaction acts as an effective magnetic field which randomly varies in time as a result of scattering processes. It results, just like in conventional Dyakonov-Perel' spin relaxation mechansim~\cite{dyakonov72,dyakonov86}, in irreversible relaxation of the exciton valley polarization and valley coherence~\cite{PSSB:PSSB201552211}. In the collision-dominated regime, the relaxation is slowed down by the scattering processes.

It is noteworthy that a magnetic field suppresses spin and valley depolarization, see Refs.~\cite{Ivchenko73,1988JETPL..47..486I,glazov04}, enhances the polarization degree of the luminescence and slows down the valley relaxation. Similar suppression of the depolarization occurs in the presence of significant spin or valley polarization of charge carriers or excitons, cf.~\cite{wu03prb,glazov04a,stich:205301,glazov_sns_pol,PhysRevB.93.241307} because an effective  magnetic field due to the exchange interaction arises in the presence of polarization. In this case, one can expect complicated and nonlinear spin and valley dynamics at high polarization with bistable behavior akin to the bistability in the interacting electron-nuclear spin system~\cite{opt_or_book,Urbaszek:2013ly,glazov2018electron}. Such bistable behavior can be strongly affected by valley polarization fluctuations which are inevitable in open driven systems~\cite{Lax1,springerlink:10.1007/BF02724353,PhysRevB.90.085303,glazov:sns:jetp16}.

Here we develop analytical theory of the valley and spin dynamics of interacting excitons in monolayer transition metal dichalcogenides in the presence of significant valley polarization of excitons. We demonstrate nonlinear temporal relaxation of polarization in these interacting conditions, bistability of the valley polarization in the steady-state regime. We also study the role of polarization fluctuations and demonstrate the noise-induced switching between the stable states.

The paper is organized as follows. Section~\ref{sec:relax} introduces the model and presents the results for the temporal dynamics (Sec.~\ref{sec:TR}) and steady states (Sec.~\ref{sec:steady}). In Sec.~\ref{sec:fluct} we study fluctuations of valley polarization in the vicinity of the stable states (Sec.~\ref{sec:small}) and switching between the steady states (Sec.~\ref{sec:switching}). \add{In Sec.~\ref{sec:disc} we discuss the routes for experimental realization of these theoretical predictions.} Brief summary of the results is presented in Sec.~\ref{sec:concl}.

\section{Exciton valley relaxation in the presence of valley polarization}\label{sec:relax}

\subsection{Kinetic equation model}\label{sec:mod}

We consider the minimal model of bright exciton doublet formed by the electronic states in the vicinity of the $\bm K_\pm$ points of the transition metal dichalcogenide Brillouin zone. We use exciton pseudospin density matrix approach and consider the dynamics of the exciton pseudospin vector $\bm s_{\bm k}$ where $\bm k$ is the exciton translational motion wavevector. The components of $\bm s_{\bm k}$ describe the distribution of the valley polarization or circular polarization of excitons, $s_{\bm k,z}$, and of the valley coherence or exciton alignment, $s_{\bm k,x}$ and $s_{\bm k_y}$. The pseudospin dynamics is governed by the kinetic equation~\cite{maialle93,PhysRevLett.81.2586,glazov2014exciton,PSSB:PSSB201552211}
\begin{equation}
\label{kinetic}
\frac{\partial \bm s_{\bm k}}{\partial \bm t} + \bm s_{\bm k} \times \bm \Omega_{\bm k} + \bm Q\{\bm s_{\bm k}\} + \frac{\bm s_{\bm k}}{\tau_0} = \bm g_{\bm k}(t),
\end{equation}
where  $\bm \Omega_{\bm k}$ is the exciton pseudospin precession frequency related to the LT-splitting of the exciton states and exciton-exciton interaction, $\bm Q\{\bm s_{\bm k}\}$ is the collision integral, $\tau_0$ is the exciton lifetime, and $\bm g_{\bm k}(t)$ is the exciton generation rate. In what follows we take the collision integral in the simplest possible form $\bm Q\{\bm s_{\bm k}\}  = (\bm s_{\bm k} - \bar{\bm s}_{\bm k})/\tau$, where $\tau$ is the scattering time. Note that $\tau$ also includes the contributions due to the exciton-exciton scattering, see Refs.~\cite{glazov02,wu03prb,glazov04a,glazov05a} for details.

The pseudospin precession frequency \add{can be interpreted as an effective magnetic field acting on the exciton pseudospin. It} contains two contributions, $\bm \Omega_{\bm k} = \bm \Omega^{LT}(\bm k) +\bm \omega(\bm S)$. The first one, $\bm \Omega^{LT}(\bm k)$ is related to the longitudinal-transverse splitting of the excitonic states caused by the long-range exchange interaction between the electron and hole in the exciton~~\cite{glazov2014exciton,Yu:2014fk-1,PhysRevB.89.205303,PhysRevB.90.161302,PSSB:PSSB201552211,prazdnichnykh2020control}. It can be represented as
\begin{equation}
\label{omega:LT}
\bm \Omega^{LT}_{\bm k} = \frac{\Delta E_{LT}(k)}{\hbar}\left[\cos{(2\varphi_{\bm k})}\hat{\bm x} + \sin{(2\varphi_{\bm k})}\hat{\bm y}\right].
\end{equation}
Here $\Delta E_{LT}(k)$ is the longitudinal-transverse splitting which scales approximately linearly with the absolute value of the wavevector~\cite{glazov2014exciton,prazdnichnykh2020control}, $\varphi_{\bm k}$ is the angle of the wavevector in the monolayer plane, $\hat{\bm x}$, $\hat{\bm y}$, and $\hat{\bm z}$ are the unit vectors along the corresponding axes, $\hat{\bm z}$ is the monolayer normal. \add{Equation~\eqref{omega:LT} describes the splitting of the excitonic states into the longitudinal one (with the microscopic dipole moment oscillating parallel to the exciton wavevector $\bm k$) and transversal one (with the microscopic dipole moment perpendicular to $\bm k$)~\cite{glazov2014exciton}. In the pseudospin approach, the longitudinal-transverse splitting acts as an effective magnetic field causing the precession of the vector $\bm s_{\bm k}$ around $\bm \Omega_{\bm k}^{LT}$. Microscopically, $\Delta E_{LT}(k)$ is related to the interaction of exciton with the induced electromagnetic field, see Refs.~\cite{glazov2014exciton,prazdnichnykh2020control,BP}. 
The same model applies to the quasi-two-dimensional excitons in conventional quantum well structures~\cite{maialle93,goupalov98,ivchenko05a}. Note that in bulk semiconductors the exciton fine structure is different and in three-dimensional systems exciton-polariton formation should also be  taken into account~\cite{BP,DM,PI}. }

The second contribution to the pseudospin precession frequency results from the exchange interaction between the excitons~\cite{Amand:1994wx,fernandez-r96,ciuti98,inoue00,bentaboudeleon01,betbeder02a,PhysRevB.80.155306}\add{: Due to the antisymmetry of the two-exciton wavefunction with respect to permutations of identical fermions the Coulomb interaction depends on the mutual orientation of exciton spins}. This contribution is nonlinear in the exciton valley imbalance and can be presented as~\cite{ivchenko05a,PhysRevB.72.075317,PhysRevB.73.073303}
\begin{equation}
\label{omega:exc}
\bm \omega(\bm S) = V_1 S_z \hat{\bm z} + V_2 (S_x \hat{\bm x} + S_y\hat{\bm y}),
\end{equation}
where $V_1$ and $V_2$ are the exchange interaction constants, $\bm S=(S_x,S_y,S_z)\equiv \sum_{\bm k} \bm s_{\bm k}$ is the total pseudospin of excitons. \add{The constant $V_1 \sim E_B a_B^2/\hbar$ with $E_B$ being the exciton binding energy and $a_B$ its Bohr radius~\cite{ciuti98,TY,PhysRevB.80.155306,Vanik}, the constant $V_2$ is typically smaller in absolute value but can be negative, see Ref.~\cite{ciuti98,PhysRevB.80.155306,Ivanov}. Equation~\eqref{omega:exc} holds for $E_B a_B^2 |S| \ll 1$, this condition is typically fulfilled even for highest possible exciton polarization for reasonable exciton densities in transition-metal dichalcogenide monolayers $n_x \lesssim 10^{11}$~cm$^{-2}$.} The presence of an effective field $\bm\omega(\bm S)$ induces an additional term to the LT splitting of excitons and suppresses the valley relaxation similarly to the Hartree-Fock effect in the spin-polarized electron gas~\cite{wu03prb,glazov04a}. To describe the effect we assume that excitons are initially polarized by circularly polarized light with the pseudospin aligned along the $z$-axis and that $\Omega^{LT}_{\bm k}\tau \ll 1$, i.e., the strong scattering regime is realized. Iterating Eq.~\eqref{kinetic} over $\Omega^{LT}_{\bm k}$ we derive the equation governing the dynamics of the total pseudospin in the ensemble as~\cite{glazov04a}
\begin{equation}
\label{Sz:gen}
\frac{dS_z}{dt} + \frac{S_z}{\tau_v(S_z)}+ \frac{S_z}{\tau_0} = G_z(t),
\end{equation}
where $G_z(t) = \sum_{\bm k} g_{\bm k,z}(t)$, and 
\begin{equation}
\label{tau:v:1}
\frac{1}{\tau_v(S_z)} = \frac{\tau_{v,0}^{-1}}{1+\alpha^2 S_z^2}, 
\end{equation}
\add{with 
\begin{equation}
\label{alpha:tauv0}
\alpha = V_1\tau, \quad \tau_{v,0}= \frac{\hbar^2}{\langle E_{LT}^2(k) \tau\rangle },
\end{equation}
and the angular brackets denoting average of the exciton ensemble, see Refs.~\cite{glazov2014exciton,PSSB:PSSB201552211,prazdnichnykh2020control}. Here $\tau_{v,0}$ is the valley relaxation time for negligible polarization of excitons, $\tau_v(S_z=0) = \tau_{v,0}$, and the parameter $\alpha$ describes the strength of exciton-exciton interaction. Physically, the presence of exciton spin polarization produces the contribution $\bm \omega(\bm S)$, Eq.~\eqref{omega:exc}, to the pseudospin precession frequency and can be interpreted as an effective magnetic field arising as a result of exciton polarization akin to the exchange field in magnetic media. As a result, the exciton pseudospin precesses in the total effective field $\bm \Omega_{\bm k}^{LT} + \bm \omega(\bm S)$ and the role of the $\bm k$-dependent longtidinal-transverse field diminishes, hence, the exciton valley relaxation becomes suppressed and $\tau_v$ increases with increase in $S_z$.} The dependence of the valley relaxation time on the exciton pseudospin, Eq.~\eqref{tau:v:1}, makes exciton valley dynamics nonlinear. \add{Noteworthy, Eqs.~\eqref{Sz:gen} and \eqref{tau:v:1} can be applied to study the exciton spin dynamics in conventional quantum wells and the resident electron spin dynamics in semiconductor quantum wells based on II-VI and III-V materials where the spin relaxation is controlled by the Dyakonov-Perel mechanism, see Refs.~\cite{wu03prb,glazov04a,stich:176401,stich:205301} for details.}

\add{We stress that Eqs.~\eqref{Sz:gen} and \eqref{tau:v:1} hold in the strong scattering regime where $\Omega^{LT}_{\bm k}\tau \ll 1$ (strong scattering) or $(\alpha S_z)^2\gg 1$ (high polarization). In atomically thin WSe$_2$ the strong scattering regime was realized, e.g., in Ref.~\cite{PhysRevB.90.161302} (see also Sec.~\ref{sec:disc} for details), and in conventional quantum well structures in, e.g., Ref.~\cite{Vinattiery} or Ref.~\cite{dyakonov_book} for review. }

In the remaining parts of this section we address non-stationary and stationary solutions of Eq.~\eqref{Sz:gen}.

\subsection{Transient dynamics of valley polarization}\label{sec:TR}

Here we consider the situation typically realized in the time-resolved experiments where the excitons are created by short circularly polarized laser pulse, see Ref.~\cite{PhysRevB.90.161302}. Accordingly, we take $G_z(t) = S_0\delta(t)$, where $\delta(t)$ is the Dirac $\delta$-function and recast the solution of eq.~\eqref{Sz:gen} in the implicit form
\begin{multline}
\ln{\left(\frac{S_z(t)}{S_0}\right)} + \frac{\tau_0}{2\tau_{v,0}} \ln{\left(\frac{\tau_0 + \tau_{v,0}[1+ \alpha^2S_z^2(t)]}{\tau_0 + \tau_{v,0}[1+ \alpha^2S_0^2(t)]}\right)}\\
= -\frac{t}{\tau_{V,0}},
\end{multline}
where 
\begin{equation}
\label{tau:V0}
\tau_{V,0} = \frac{\tau_0 \tau_{v,0}}{\tau_0+ \tau_{v,0}}.
\end{equation}

There are two parameters which control the valley dynamics. The first one is the ratio $\tau_0/\tau_{v,0}$, it characterizes the ratio of the exciton lifetime and valley relaxation time. The second parameter is $|\alpha S_0|$, it controls the strength of the nonlinearity, i.e., the feedback of polarized excitons on their valley relaxation. 

\begin{figure}[t]
\includegraphics[width=0.99\linewidth]{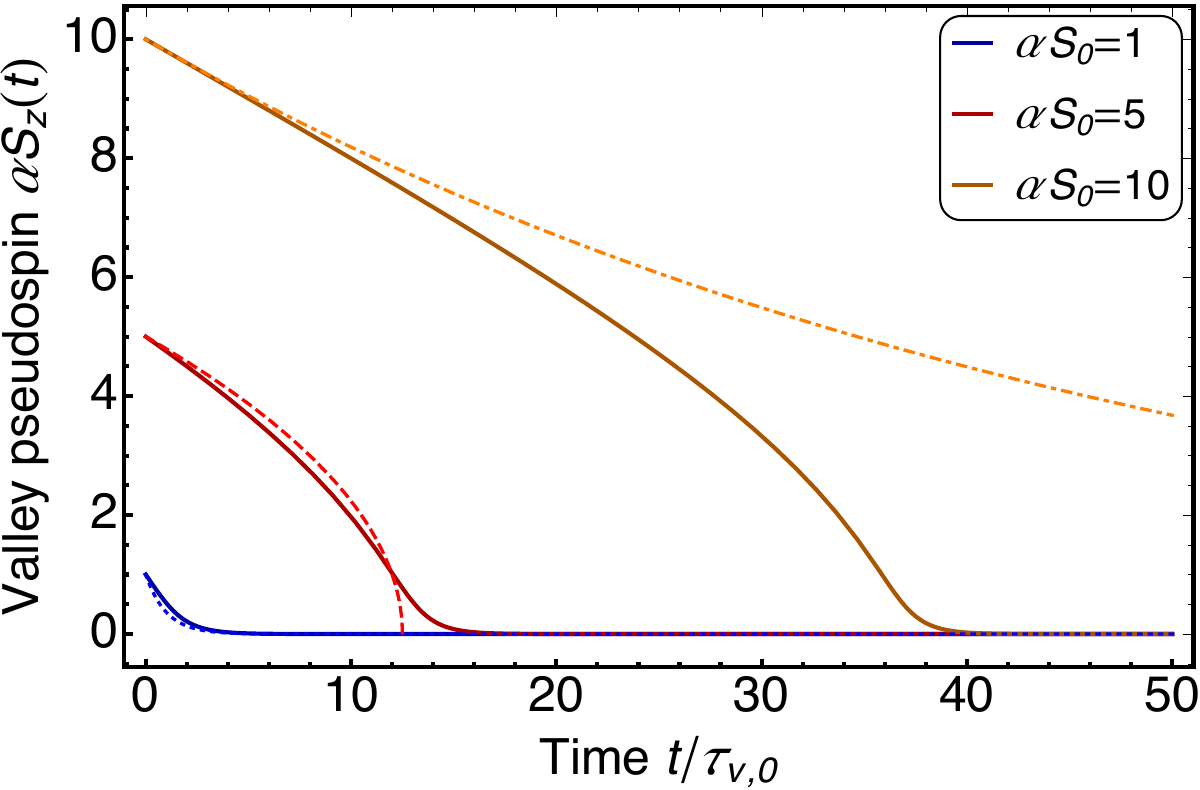}
\caption{Temporal dependence of the $z$-component of the pseudospin calculated at different initial valley polarizations: $\alpha S_0=1$ (blue), $5$ (red), and $10$ (orange). Dotted, dashed, and dash-dotted lines show analytical asymptotes Eqs.~\eqref{small}, \eqref{moderate:0}, and \eqref{high}, respectively.}\label{fig:time}
\end{figure}

If $\tau_0 \ll \tau_{v,0}$ then exciton lifetime is so short that the valley relaxation does not occur during the lifetime of exciton. Therefore, in this regime $S_z(t) =S_0(t) \exp{(-t/\tau_0)}$. Here and in what follows we focus on the opposite limit where $\tau_0 \gg \tau_{v,0}$. In this case, for small polarization $|\alpha S_z| \ll 1$ valley relaxation is also exponential and controlled by $\tau_{v,0}$: 
\begin{equation}
\label{small}
S_z(t) =S_0(t) \exp{(-t/\tau_{v,0})}.
\end{equation} 
The most interesting situation occurs, however, where $|\alpha S_0|$ is significant. For moderate polarization, where $|\alpha S_0| \gg 1$, but $\alpha^2 S_0^2 \ll \tau_{0}/\tau_{v,0}$ we have
\begin{equation}
\label{moderate:0}
S_z(t) = S_0 \sqrt{1-\frac{2t}{\alpha^2S_0^2\tau_{v,0}}}.
\end{equation}
Such a dependence holds as long as $|\alpha S_z|\gg 1$ and transforms to $\exp{(-t/\tau_{v,0})}$ at longer timescales. Effective magnetic field \add{$\bm \omega(\bm S)$} induced by the exciton polarization stabilizes pseudospin and prevents its relaxation.

At higher polarizations, where $\alpha^2 S_0^2 \gg \tau_{0}/\tau_{v,0}$, the relaxation is, at first, exponential and controlled by the effective valley lifetime $\tau_V$,
\begin{equation}
\label{high}
S_z =S_0 \exp{(-t/\tau_V)}, \quad \frac{1}{\tau_V} = \frac{1}{\tau_0}  + \frac{1}{\tau_v(S_0)},
\end{equation}
 until $\alpha^2 S_z^2$ becomes comparable with $\tau_{0}/\tau_{v,0}$. Then the polarization decay is given by Eq.~\eqref{moderate:0}. 

Time dependence of the exciton pseudospin is shown in Fig.~\ref{fig:time} for different initial conditions illustrating the analytical model developed above. Generally, an increase of the  valley polarization results (i) in a non-exponential valley relaxation and (ii) in an enhancement of the valley lifetime. \add{Similar effects were observed for electron spin relaxation in GaAs quantum wells~\cite{stich:176401,stich:205301}.}

\subsection{Steady-state valley polarization}\label{sec:steady}

\begin{figure}[t]
\includegraphics[width=0.99\linewidth]{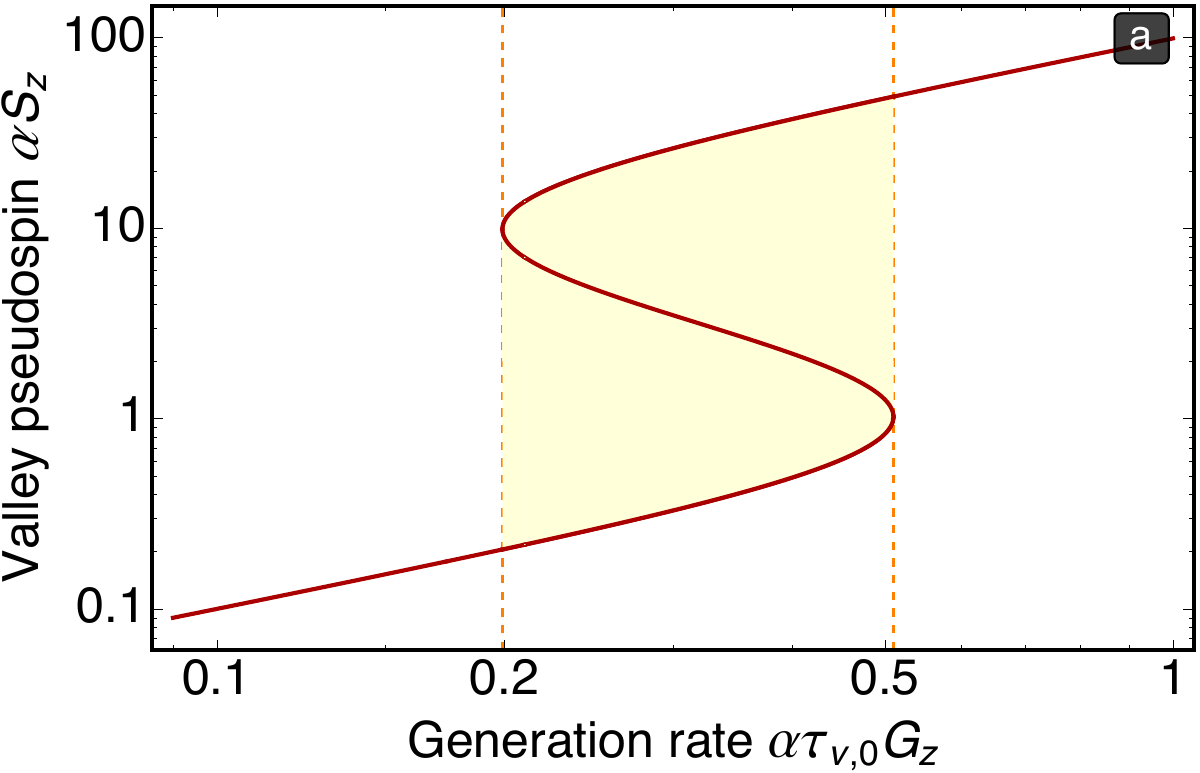}\\
\vspace{0.25cm}
\includegraphics[width=0.99\linewidth]{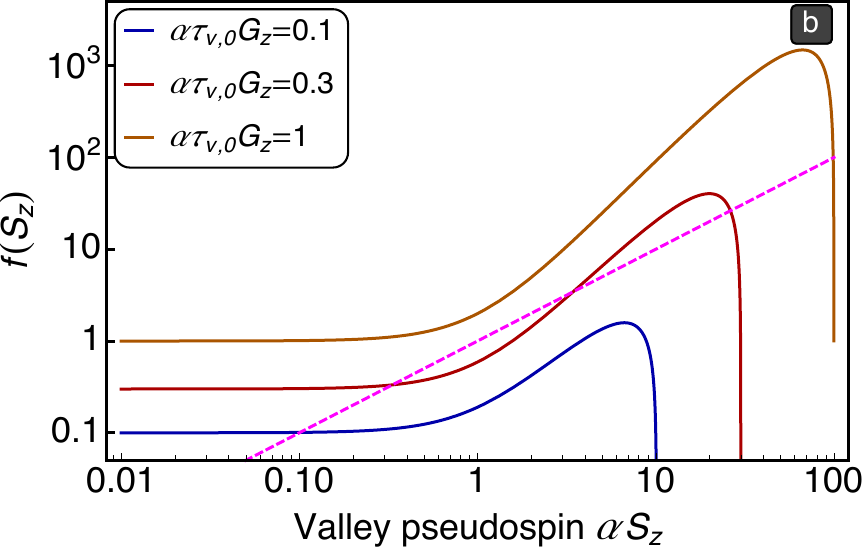}\\
\caption{(a) Valley pseudospin as function of the generation rate calculated after solution of Eq.~\eqref{Sz:work}. The area with bistable behavior is shown by the light yellow shading. (b) Graphical illustration of Eq.~\eqref{Sz:work:1}. Solid curves show the left-hand side, $f(S_z)$, for different values of the generation rate $\alpha\tau_{v,0} G_z=0.1$ (blue), $0.3$ (dark red), $1$ (orange). Dashed magenta curve shows the right hand side, $S_z$. $\tau_0/\tau_{v,0}=100$.}\label{fig:steady}
\end{figure}

Let us now switch to the steady-state situation where the excitons are generated by a \emph{cw} source. In this case, $G_z$ does not depend on the time and time-derivative in Eq.~\eqref{Sz:gen} can be omitted. Here we neglect pumping fluctuations and related fluctuations of the valley pseudospin; the effects of noise are discussed in detail below in Sec.~\ref{sec:fluct}. Thus, Eq.~\eqref{Sz:gen} transforms to the following cubic equation 
\begin{equation}
\label{Sz:work}
-\frac{\alpha^2}{\tau_0} S_z^3 + G_z\alpha^2 S_z^2 - \frac{S_z}{\tau_{V,0}} + G_z=0.
\end{equation}
Figure~\ref{fig:steady}(a) shows the dependence of the valley pseudospin $S_z$ found from Eq. \eqref{Sz:work} as a function of the generation rate $G_z$ with the region with several solutions is clearly visible.

While general analytical and numerical solutions of Eq.~\eqref{Sz:work} are readily available, to get better insight in the physics of the effect it is convenient to study the solutions of Eq.~\eqref{Sz:work} in the graphical form, representing it as
\begin{equation}
\label{Sz:work:1}
f(S_z) = S_z, \quad f(S_z) = -\frac{\alpha^2 \tau_{v,0}}{\tau_0} S_z^3 + G_z\tau_{v,0} \alpha^2 S_z^2 + G_z\tau_{v,0}.
\end{equation}
Here we omitted the term with $S_z/\tau_0$ assuming that $\tau_0 \gg \tau_{v,0} \approx \tau_{V,0}$, see Eq.~\eqref{tau:V0}.
Left-hand and right-hand sides of Eqs.~\eqref{Sz:work:1} are plotted in Fig.~\ref{fig:steady}(b). Accordingly, three regimes depending on the number and positions of roots of Eq.~\eqref{Sz:work:1} are clearly seen.

The first regime is a \emph{weak pumping regime}, blue curve in Fig.~\ref{fig:steady}(b), where non-linearity is umimportant and Eq.~\eqref{Sz:work:1} has only one solution 
\begin{equation}
\label{Sz:weak}
S_z \approx G_z\tau_{v,0}.
\end{equation}
This regime is realized at
\begin{equation}
\label{weak}
G_z \lesssim \frac{3}{\sqrt{2\alpha^2\tau_0\tau_{v,0}}}.
\end{equation}

The second regime is a \emph{moderate pumping regime} where
\begin{equation}
\label{moderate}
\frac{2}{\sqrt{\alpha^2\tau_0\tau_{v,0}}}\lesssim G_z   \lesssim  \frac{1}{2\alpha \tau_{v,0}}, 
\end{equation}
and corresponds to the red curve in Fig.~\ref{fig:steady}(b). In this situation Eq~\eqref{Sz:work:1} has three solutions, $S^{(1)}_z < S^{(2)}_z < S^{(3)}_z$:
\begin{subequations}
\begin{align}
\label{Sz:moderate}
&S_z^{(1)} \approx G_z \tau_{v,0},\\
&S_z^{(2)} \approx 1/(G_z\tau_{v,0}\alpha^2), \\
&S_z^{(3)} \approx G_z\tau_0.
\end{align}
\end{subequations}
We show below that the moderate pumping regime corresponds to the bistable behavior of the valley polarization.

Finally, a \emph{strong pumping regime} is realized provided 
\begin{equation}
\label{strong}
\alpha G_z \tau_{v,0}  \gg 1, 
\end{equation}
and in this case, Eq.~\eqref{Sz:work:1} has only one solution
\begin{equation}
\label{Sz:strong}
S_z \approx G_z\tau_0.
\end{equation}

Let us analyze the moderate pumping regime in more details. First of all, we address the stability of solutions $S^{(i)}$, $i=1,2,3$, presented in Eqs.~\eqref{Sz:moderate}. We evaluate the temporal dynamics of small fluctuations $\delta S_z^{(i)} = S_z - S_z^{(i)}$ in the vicinity of the $i$th solution by linearizing Eq.~\eqref{Sz:gen} with the result:
\begin{equation}
\label{fluct}
\frac{d \delta S_z^{(i)}}{dt} + \lambda_i \delta S_z^{(i)} =0,
\end{equation}
where 
\begin{equation}
\label{damping}
\lambda_i =\frac{1}{\tau_0} + \frac{1}{\tau_{v,0}} \frac{1-\left(\alpha S_z^{(i)}\right)^2}{\left[1+\left(\alpha S_z^{(i)}\right)^2\right]^2}.
\end{equation}
We analyze dynamics of spin fluctuations in more detail in the following Sec.~\ref{sec:fluct}. Here we check that $\lambda_{1,3}>0$, while $\lambda_2<0$. Hence, the solutions $S_z^{(1)}$ and $S_z^{(3)}$ are stable, while the solution $S_z^{(2)}$ is unstable. This result is similar to the case of the classical anharmonic oscillator~\cite{ll1_eng} which also demonstrates $S$-like response, cf. Fig.~\ref{fig:steady}(a), with high- and low-amplitude solutions being stable and the intermediate one being unstable.

\begin{figure}[h]
\includegraphics[width=0.99\linewidth]{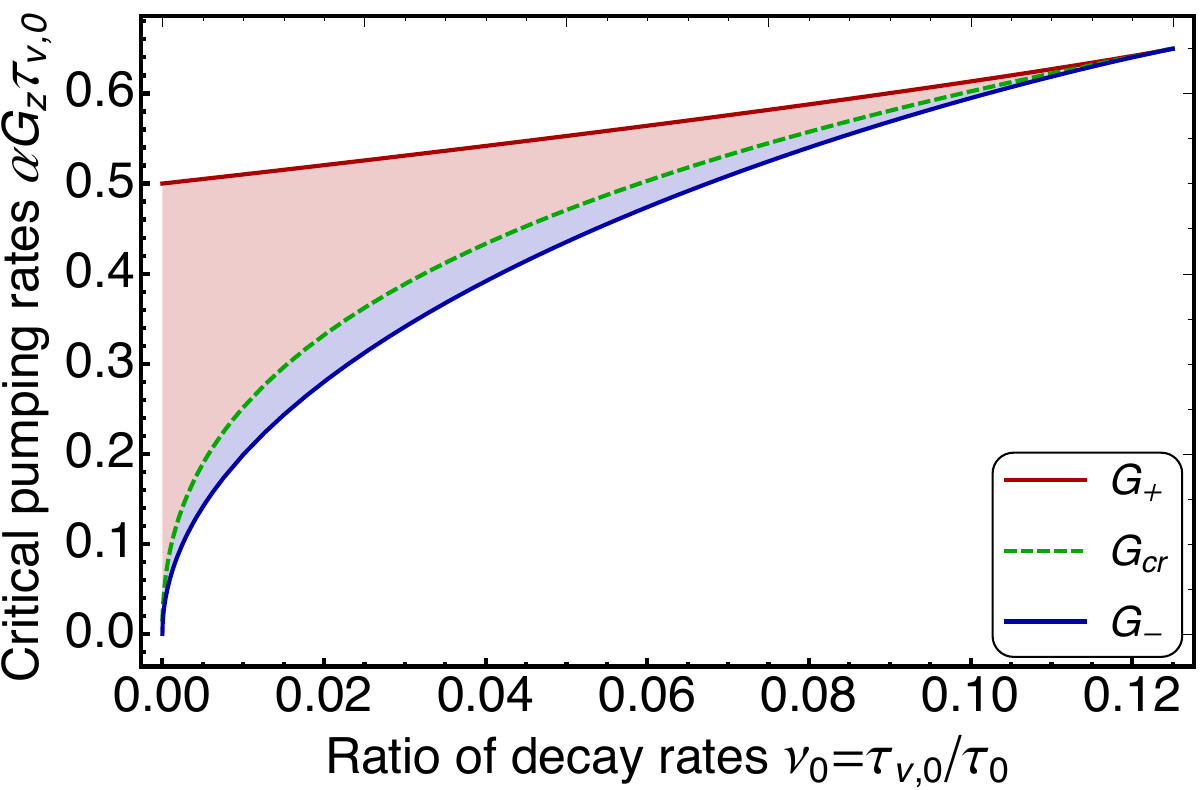}
\caption{Range of pumping rates where the bistable behavior is possible. Solid red and blue curves show upper and lower bounds for $G_z$, see Eqs.~\eqref{moderate:1} and \eqref{Gpm}. Dashed green curve shows critical transition rate $G_{cr}$, Eq.~\eqref{G:cr}, corresponding to the predominant realization of the $S_z^{(1)}$ or $S_z^{(3)}$ in the case of the noisy pumping. }\label{fig:crit}
\end{figure}

Now let us specify the conditions for the bistable behavior to take place. We introduce the parameter $\nu_0=\tau_{v,0}/\tau_0$ characterizing the ratio between the valley relaxation time and the lifetime of excitons. The analysis of the cubic equation~\eqref{Sz:work} shows that the bistability is possible for $\nu_0<1/8$. The bistability region is realized for the range of the generation rates
\begin{equation}
\label{moderate:1}
G_-<G_z<G_+,
\end{equation}
where
\begin{subequations}
\label{Gpm}
\begin{align}
\alpha \tau_{v,0}G_- = \frac{\sqrt{1+20\nu_0-8\nu_0^2-(1-8\nu_0)^{3/2}}}{2\sqrt{2}},\\
\alpha \tau_{v,0}G_+ = \frac{\sqrt{1+20\nu_0-8\nu_0^2+(1-8\nu_0)^{3/2}}}{2\sqrt{2}}.
\end{align}
\end{subequations}
For $\nu_0 \ll 1$ Eq.~\eqref{moderate:1} passes to Eq.~\eqref{moderate}. The boundaries for the bistable behavior are shown by red and blue lines in Fig.~\ref{fig:crit}.

\section{Fluctuations and switching between stable steady-states}\label{sec:fluct}

We now turn to the analysis of fluctuations in the valley-polarized excitonic system. \add{On the one hand, random events of the photon absorption, intrinsic fluctuations in the pumping intensity, and randomness in the scattering processes of excitons results in fluctuations in the generation rate~\cite{Lax1,Kogan,springerlink:10.1007/BF02724353,RevModPhys.68.801} with $G_z(t)$ in Eq.~\eqref{Sz:gen}. On the other hand, the spin noise experimental technique~\cite{AZ,Oestreich} is now actively applied to non-equilibrium systems~\cite{PhysRevB.93.241307,Crooker} and to two-dimensional materials to detect the valley fluctuations~\cite{Crooker:2d}. To account for the fluctuations we take $G_z(t)$ in the form}
\begin{equation}
G_z(t) = \langle G_z\rangle + \delta G(t).
\end{equation}
Here $\langle G_z\rangle$ is the time-average value of the pumping rate and $\delta G_z(t)$ is its random fluctuation. The \add{fluctuations or} noise of the pumping is characterized by autocorrelation function which we take in the simplest form
\begin{equation}
\label{dG:fluct}
\langle \delta G(t) \delta G(t') \rangle = 2\beta \delta(t-t'),
\end{equation}
where the constant $\beta$ describes the intensity of fluctuations. \add{Since the characteristic time-scale of fluctuations is the exciton scattering time $\tau$ or even shorter and the exciton valley dynamics occurs on a longer timescale, $\tau_v(S_z) \gg \tau$, see Eq.~\eqref{tau:v:1}, the white-noise assumption in Eq.~\eqref{dG:fluct} is well justified.} Note, that we use classical equations to describe the valley noise of excitons, accordingly, we assume that noise $\delta G$ includes also random Langevin forces which support quantum statistical fluctuations of the particles and spins~\cite{ll5_eng,aronovivchenko71:eng,gi2012noise,smirnov:SNS:rev}. 

\begin{figure}[t]
\includegraphics[width=0.99\linewidth]{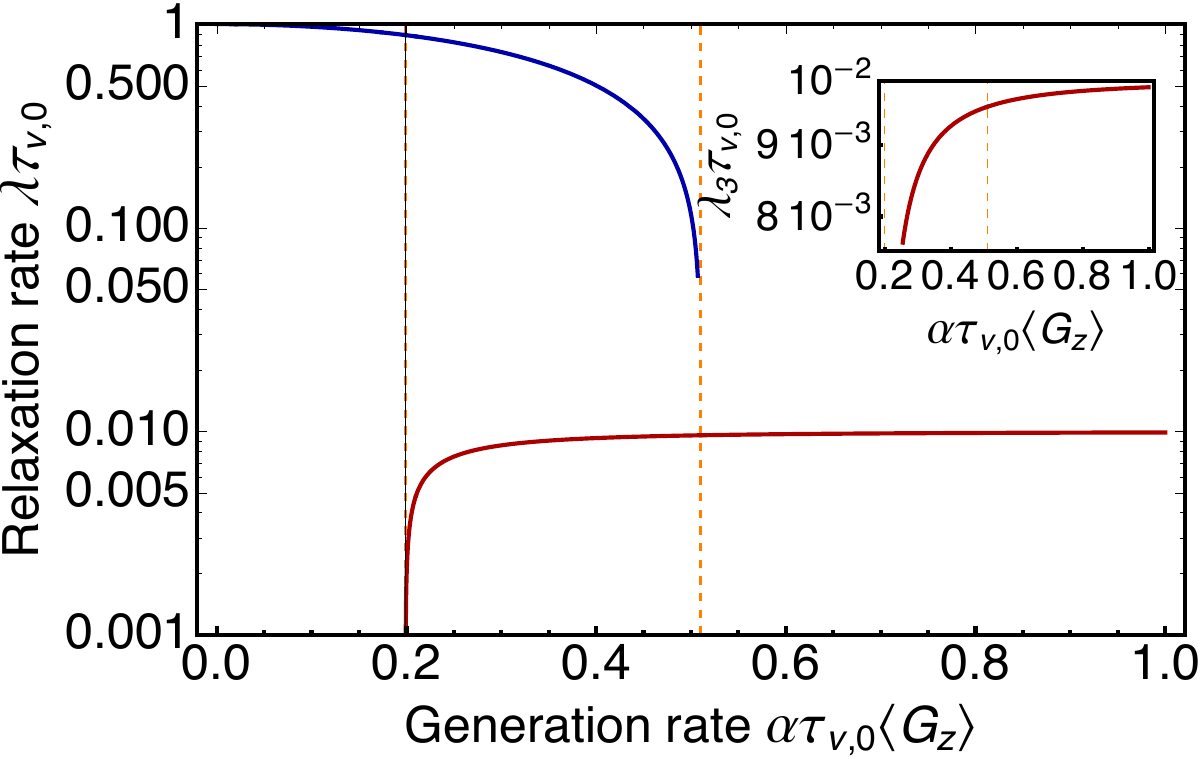}
\caption{Relaxation rates of small fluctuations in the vicinity of the steady-state solutions, Eq.~\eqref{damping}. Blue curve shows $\lambda_1$ corresponding to the low valley imbalance, $S_z^{(1)}$, red curve shows $\lambda_3$ corresponding to the hight alley imbalance, $S_z^{(3)}$. Inset shows $\lambda_3$ in a linear scale. $\tau_0/\tau_{v,0}=100$.  }\label{fig:lambda}
\end{figure}

\begin{figure*}[t]
\includegraphics[width=0.7\linewidth]{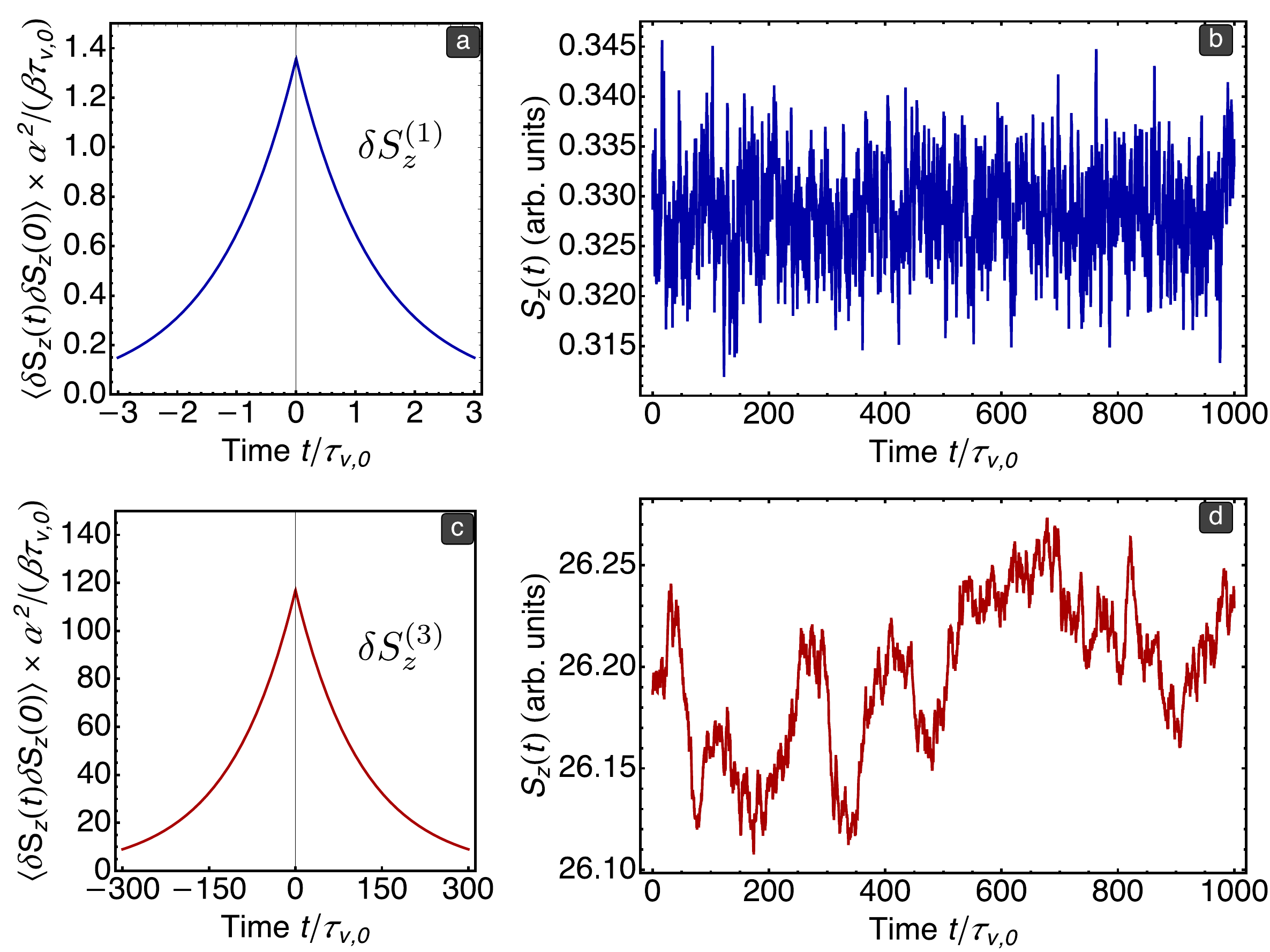}
\caption{Pseudospin autocorrelation functions, Eq.~\eqref{noise:small} calculated for the steady state solutions with low valley imbalance, $S_z^{(1)}$ (a), and high valley imbalance, $S_z^{(3)}$ (c). Corresponding fluctuating time-dependent traces $S_z(t)$ found by direct numerical solution of Eq.~\eqref{Sz:gen} with fluctuating $G_z(t)$ according to Eq.~\eqref{dG:fluct}. $\alpha \langle G_z(t)\rangle \tau_{v,0}=0.3$, $\beta=3\times 10^{-4}/(\alpha^2\tau_{v,0})$, $\tau_0/\tau_{v,0}=100$.  }\label{fig:noise}
\end{figure*}

\subsection{Small fluctuations}\label{sec:small}

In the limit of $\beta\to 0$ one can readily calculate the correlation function of pseudospin $z$-components using linearlized version of Eq.~\eqref{Sz:gen} [cf. Eqs.~\eqref{fluct}]
\begin{equation}
\label{fluct:1}
\frac{d \delta S_z^{(i)}}{dt} + \lambda_i \delta S_z^{(i)} =\delta G(t).
\end{equation}
We recall that $\delta S_z^{(i)} = S_z(t) - S_z^{(i)}$ with $S_z^{(i)}$ ($i=1,2,3$) being the steady-state solution discussed in previous Sec.~\ref{sec:steady}, and $\lambda_i$ is the relaxation rate presented in Eq.~\eqref{damping}. Following Refs.~\cite{ll5_eng,glazov2018electron,glazov:sns:jetp16} we obtain
\begin{equation}
\label{noise:small}
\langle \delta S_{z}^{(i)}(t) \delta S_{z}^{(i)}(t')\rangle = \frac{\beta}{\lambda_i} e^{-\lambda_i|t-t'|},
\end{equation}
so the valley noise spectrum has a Lorentzian form with the half-width at half-maximum being $1/\lambda_i$.

The dependence of the relaxation rates $\lambda_1$ and $\lambda_3$ is shown in Fig.~\ref{fig:lambda}. Note that $\lambda_2<0$, intermediate state is unstable, corresponding fluctuations grow linearly in time and present linear analysis is invalid, see Sec.~\ref{sec:switching} for details. In the case of weak pumping one has only one solution $S_z^{(1)}$ with relatively large relaxation rate $\lambda_1 \approx 1/\tau_{v,0}$, see blue curve in Fig.~\ref{fig:lambda}, which decreases with increasing the $\langle G_z\rangle$. In the high pumping regime we have again only one solution $S_z^{(3)}$ with relatively small damping rate $\lambda_3 \approx 1/\tau_0 \ll \lambda_1$. In the bistable region, two solutions $S_z^{(1)}$ and $S_z^{(3)}$ coexist with $\lambda_1> \lambda_3$.

Figure~\ref{fig:noise} shows pseudospin autocorrelation functions and typical examples of $S_z$ time-dependence. Here we focus on the bistability regime. Panels (a,b) correspond to the steady-state solution with low valley imbalance $S_z^{(1)}$. Corresponding fluctuations are relatively weak and characterized by the short correlation time. For the high valley imbalance solution is presented in panels (c,d). The fluctuations in that regime are larger and their correlation time is much longer than in low valley imbalance regime. This is in agreement with Eq.~\eqref{noise:small} which clearly shows that the smaller is the relaxation rate the larger fluctuations accumulate in the system.

\subsection{Switching between the stable states}\label{sec:switching}

\begin{figure}[t]
\includegraphics[width=0.99\linewidth]{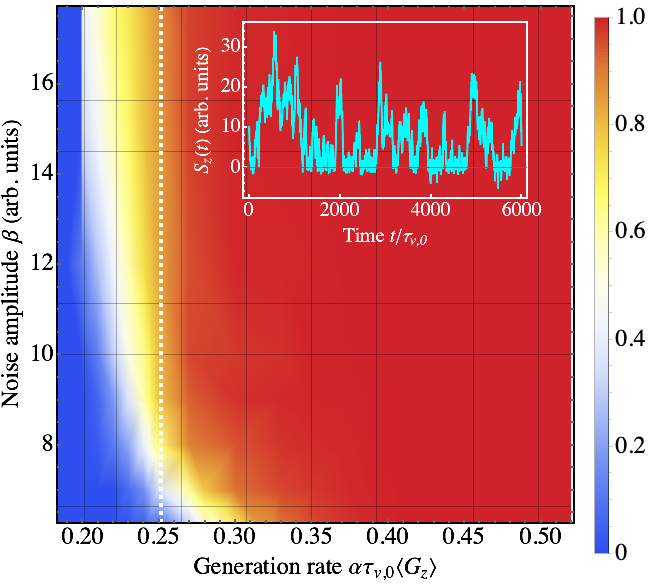}
\caption{Fraction of time when the high $S_z$ solution ($\langle S_z \rangle = S_z^{(3)}$) is realized. Blue color shows the stability area of the low-$S_z$ solution ($\langle S_z \rangle = S_z^{(1)}$), red color shows the high $S_z$ solution ($\langle S_z \rangle = S_z^{(3)}$) is realized. Vertical dashed line shows $G_{cr}$ found from Eq.~\eqref{G:cr} condition.  Inset shows typical time dependence of $S_z(t)$ at $\alpha\tau_{v,0} \langle G \rangle =0.21$ and sufficiently high noise amplitudes where switching between the high-$S_z$ and low-$S_z$ regimes is observed. $\tau_0/\tau_{v,0}=100$.}\label{fig:switch}
\end{figure}

Fluctuations of the pumping, $\delta G(t)$, and induced fluctuations $\delta S_z(t)$ beyond the linear approximation Eq.~\eqref{fluct:1} can significantly affect the steady-state of the system and, in particular, result in switching between the steady-states $1$ and $3$. Indeed, sufficiently large fluctuations of pumping $\delta G$ can bring the system to the vicinity of the unstable solution $S_z \approx S_z^{(2)}$. Then, the initial state will be essentially ``forgotten'': even a negligibly small fluctuation exponentially increases $\propto \exp{(|\lambda_2| t)}$, and renders the system to one of the stable states with almost the same probability regardless the initial state of the system.

Let us study the switching effect in more detail. To that end, we follow Refs.~\cite{Ventsel__1970,dykman79} (see also Refs.~\cite{dmitriev,maslova,Maslova:2009tg} for alternative approaches) and represent the probability density $w(t_1,S_{z,1} \to t_2, S_{z,2})$ for realizing the classical `trajectory' in the $(t,S_z)$ space of Eq.~\eqref{Sz:gen} that starts at $t=t_1$ from $S_{z,1}$ and reaches at $t=t_2$ the point $S_{z,2}$. We assume that $S_{z,1}$ is close to one of the stable steady-states, $S_{z,1} \approx S_z^{(1)}$ or $S_z^{(3)}$, while $S_{z,2}$ is close to the unstable intermediate state $S_z^{(2)}$. Any trajectory $S_z(t)$ corresponds to a realization of pumping fluctuation $\delta G(t)$ which can be readily found as from Eq.~\eqref{Sz:gen}
\[
\delta G(t) = \frac{dS_z}{dt} + \frac{S_z}{\tau_v(S_z)}+ \frac{S_z}{\tau_0}  - \langle G_z \rangle.
\]
The probability distribution for a white-noise function $\delta G(t)$ has the form
\begin{equation}
\label{prop:dG}
\mathcal P[\delta G(t)]\propto \exp{\left(-\frac{1}{2\beta} \int |\delta G(t)|^2 dt \right)}.
\end{equation}
Hence, within the exponential accuracy one can write the transition probability as
\begin{equation}
\label{trans}
w(t_1,S_{z,1} \to t_2, S_{z,2}) \propto \max\{\exp{(-S/2\beta)}\}, 
\end{equation}
where
\[
\quad S = \int_{t'}^{t_2} L[S_z,\dot{S}_z] dt,
\]
$t'$ is the time when the fluctuation which brings the system close to the unstable solution starts, and the maximum in Eq.~\eqref{trans} is found over all trajectories $[S_z(t),\dot{S}_z(t)]$ with the dot on top denoting the time-derivative, and over $t'-t_2$ under condition that $S_z(t_2) = S_{z,2}$ and $S_z(t') \approx S_{z,1}$. 
The quantities $S$ and $L$ can be considered as the effective action and Lagrange function of a dynamical system, and
\begin{equation}
\label{lagrange}
L[S_z,\dot{S}_z] = \left[\dot{S}_z - v(S_z)\right]^2,
\end{equation}
with $v(S_z) =  \langle G_z\rangle - S_z/\tau_v(S_z) - S_z/\tau_0$.

Let us now evaluate the probability for system to switch from the stable steady-state solution $S_z^{(j)}$ ($j=1$ or $3$) and arrive at the stable steady-state solution $\bar j$, where $\bar j=3$ for $j=1$ and vice versa. This transition probability between the stable solutions can be estimated as the probability to reach an unstable one $S_z \approx S_z^{(2)}$: Once the system reaches an unstable solution it goes to another stable point with the probability close to $1/2$. Hence, 
\begin{equation}
\label{trans:prob}
W_{\bar j \leftarrow j}(t) \propto \exp{(-Q_j/2\beta)}, 
\end{equation}
where $Q = \min\int_0^t \left[\dot S_z - v(S_z)\right]^2 dt$. The minimum is taken over all possible trajectories  with $S_z(0)\approx S_z^{(j)}$ and $S_z(t) \approx S_z^{(2)}$, and time $t$. The latter can be found using standard Euler-Lagrange formalism and in agreement with Refs.~\cite{ll1_eng,dykman79,Ventsel__1970} we obtain
\begin{widetext}
\begin{equation}
\label{Q:optimal}
Q_j= 4\left|\int_{S_z^{(j)}}^{S_z^{(2)}}v(S_z) dS_z\right|
=4\left|\langle G_z\rangle\left(S_z^{(2)} - S_z^{(j)} \right) 
- \frac{1}{2\tau_0} \left(S_z^{(2)} - S_z^{(j)} \right)^2 
-\frac{1}{2\tau_{v,0}}\ln\frac{1+ \alpha^2 \left(S_z^{(2)}\right)^2}{1+ \alpha^2 \left(S_z^{(j)}\right)^2}\right|.
\end{equation}
\end{widetext}

Equations~\eqref{trans:prob} and \eqref{Q:optimal} show that noise in the system results in the switching between the states with low and high valley polarization of excitons. Naturally, for small fluctuations of the pump, $\beta\to 0$, the switching is suppressed and corresponding switching time in Eq.~\eqref{trans:prob} is exponentially long. The noisier pump is, the faster are the switchings between the steady-states. 

Importantly, for arbitrary generation rate $\langle G_z\rangle$ satisfying Eq.~\eqref{moderate:1} the transition rates between stable states are not equal
\begin{equation}
\label{non:eq:rates}
W_{3\leftarrow 1} \ne W_{1\leftarrow 3}.
\end{equation}
This effect is connected to the fact, that the unstable state 2 is ``closer'' to one of the stable states: $S^{(2)} - S^{(1)} \ne S^{(3)} - S^{(2)}$. Condition~\eqref{non:eq:rates} means that the system predominantly remains in the steady state which is harder to leave. Making use of Eq.~\eqref{trans:prob} we observe that the state $S_z^{(1)}$ with low polarization is mainly realized provided that $Q_1>Q_3$, otherwise the state $S_z^{(3)}$ is predominantly realized. Corresponding critical generation rate, $\langle G_z\rangle = G_{cr}$, found from the condition
\begin{equation}
\label{G:cr}
Q_1 = Q_3,
\end{equation}
is plotted in Fig.~\ref{fig:crit} by the green dashed line. The analysis shows that for $G_-<\langle G_z\rangle < G_{cr}$ the low-$S_z$ ($\langle S_z \rangle = S_z^{(1)}$) steady state solution is mainly realized while for $G_{cr} < \langle G_z\rangle < G_+$ the solution with high $S_z$  ($\langle S_z \rangle = S_z^{(3)}$) is mainly realized.

Analytical results obtained above are corroborated by our numerical simulations shown in Fig.~\ref{fig:switch}. Figure shows the fraction of the time where, in the presence of noisy pump, the high-$S_z$ solution is realized. Correspondingly, red part of the plot shows the pumping regime where $S_z^{(3)}$ steady-state solution is realized, while the blue part of the plot shows the regime where $S_z^{(1)}$ is realized. Dashed white vertical line shows the analytical criterion $G_{cr}$ found from Eq.~\eqref{G:cr}.  A good agreement between the results of simulations and analytics is observed. The differences at low noise amplitudes are mainly related to the inaccuracies in the numerical simulations at small noise amplitudes where the switching time is particularly long.

\add{One can see that the area where the high-$S_z$ solution is realized is larger than the area with the low-$S_z$ solution, i.e., $G_+ - G_{cr} > G_{cr} - G_-$ for our set of parameters.} Comparing Figs.~\ref{fig:crit} and \ref{fig:switch} we conclude that the pumping rates range where the low-$S_z$ regime is realized is \add{generally} narrower than the range where the high-$S_z$ regime is realized: the system prefers the high-polarization regime with slow relaxation and large fluctuations.

\add{\section{Discussion and routes for experimental realization}\label{sec:disc}}

\add{We have shown that significant valley polarization of excitons in atomically-thin semiconductors (or significant spin polarization of excitons in conventional quantum wells) can result in (i) slow-down of valley relaxation and non-exponential valley polarization dynamics, Fig.~\ref{fig:time}, in time-resolved experiments, (ii) bistable behavior of the valley polarization, Fig.~\ref{fig:steady}(a) under cw excitation, (iii) slow switching between the stable states, Fig.~\ref{fig:switch}, due to the pump fluctuations. }

\add{The non-exponential time dynamics, i.e., the effect predicted in Fig.~\ref{fig:time}, could probably be evidenced in a WSe$_2$ monolayer by perfoming time and polarization resolved photoluminescence spectroscopy or pump-probe spectroscopy. Under conditions of relatively weak excitation the exciton valley dynamics has been experimentally studied in Ref.~\cite{PhysRevB.90.161302}. The experiments could  also be performed using ellipticaly-polarized pulses. In this case one can independently vary the exciton density and degree of polarization making it possible to distinguish the effects of the high-polarization suggested here from the polarization relaxation slow-down owing to the exciton-exciton scattering.}

\add{The bistability effect requires to have short spin/valley relaxation time compared to the lifetime. A good candidate in this regard could be MoSe$_2$ monolayer, well known for its very low cw photoluminescence polarization degree, a consequence of very short spin/valley relaxation time. It could be probably due to the efffect described in Ref.~\cite{PhysRevB.101.115307}.
Another possibility could be to perform the measurements in WSe$_2$ at temperatures larger than $4$ K, yielding shorter spin/valley relaxation times as shown in \cite{PhysRevB.90.161302}. For instance, at $T\gtrsim 30$ K, Ref.~\cite{PhysRevB.90.161302} reported observation of a clear decrease of the exciton valley polarization decay time $\tau_v$ down to $1.5$ ps at $T = 125$ K with almost temperature independent decay time of the photoinduced reflectivity $\tau_0 \gtrsim 10$~ps  which allows one to reach the condition $\tau_v \ll \tau_0$. In this case cw experiments based on polarization-resolved photoluminescence or Kerr rotation could be performed. The spin noise techniques could also be used in this case making it possible to reveal predicted fluctuation spectra, Fig.~\ref{fig:noise}, and switching between the stable steady states, Fig.~\ref{fig:switch}.}

\add{Another possibility to measure the predicted nonlinear phenomena is to embed the monolayers into microcavities, see Refs.~\cite{cavity1,cavity2} for review. In such systems polarization multistability can be realized~\cite{multi1,multi2} and stochastic resonance technique can be applied to study the switching between the steady states~\cite{multi3}. }

\section{Conclusion}\label{sec:concl}

To conclude, we have studied the effect of significant valley polarization of excitons on their valley dynamics \add{in atomically thin transition metal dichalcogenides}. We have demonstrated the slow-down of valley relaxation and bistable behavior of the valley imbalance as a consequence of exciton-exciton interactions.
 We have studied  the role of fluctuations in such system and shown that noise in the system results in a switching between otherwise stable steady states. The developed model can be applied, besides two-dimensional semiconductors, to excitons and electrons in conventional quantum well structures where spin or valley polarization of the quasiparticles due to the exchange interaction slows down their spin and valley relaxation. \add{Similar effects are also possible for the resident electron spin polarization in conventional quantum well structures where the electron spin dynamics is governed by the Dyakonov-Perel' spin relaxation mechanism and can be slowed down by the electron-electron exchange interaction.}

\acknowledgments

We are grateful to Fabian Cadiz for useful discussions and Anton Prazdnichnykh for help at the initial stages of the work. Analytical theory by M.M.G. was supported by RSF project 19-12-00051. Numerical simulations by M.A.S. were supported by RFBR-CNRS joint project 20-52-16303. Part of this work performed by the French group was supported by the French Agence Nationale de la Recherche under the program ESR/EquipEx+ (grant number ANR-21-ESRE-0025) and the ANR projects ATOEMS, Sizmo- 2D and Magicvalley.


%

\end{document}